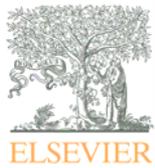

# 3D acoustic imaging applied to the Baikal Neutrino Telescope

K.G.Kebkal,[a,*] R.Bannasch,[a] O.G.Kebkal,[a] A.I.Panfilov,[b] and R.Wischnewski[c]

a) *EvoLogics GmbH, Blumenstraße 49, 10243 Berlin*
b) *Institute for Nuclear Research, 60th October Anniversary pr. 7a, Moscow 117312, Russia*
c) *DESY, Platanenallee 6, 15735 Zeuthen, Germany*

**Elsevier use only:** Received date here; revised date here; accepted date here

### Abstract

A hydro-acoustic imaging system was tested in a pilot study on distant localization of elements of the Baikal underwater neutrino telescope. For this innovative approach, based on broad band acoustic echo signals and strictly avoiding any active acoustic elements on the telescope, the imaging system was temporarily installed just below the ice surface, while the telescope stayed in its standard position at 1100 m depth. The system comprised an antenna with four acoustic projectors positioned at the corners of a 50 meter square; acoustic pulses were "linear sweep-spread signals" - multiple-modulated wide-band signals (10 ➔ 22 kHz) of 51.2 s duration. Three large objects (two string buoys and the central electronics module) were localized by the 3D acoustic imaging, with an accuracy of ~0.2 m (along the beam) and ~1.0 m (transverse). We discuss signal forms and parameters necessary for improved 3D acoustic imaging of the telescope, and suggest a layout of a possible stationary bottom based 3D imaging setup. The presented technique may be of interest for neutrino telescopes of km$^3$-scale and beyond, as a flexible temporary or as a stationary tool to localize basic telescope elements, while these are completely passive.

## 1. Introduction

Precise knowledge of the relative positions of all light sensors of an underwater neutrino telescope is essential for spatial reconstruction of particle tracks; their absolute geo-referenced positions are needed to point back to astronomical sources.

Usually, this calibration is done by an "acoustic positioning system", made up by a number of telescope elements equipped with active acoustic

─────
* Corresponding author. +49-30-31472658; fax: +49-30-31472658; e-mail: kebkal@evologics.de

beacons and a geo-referenced antenna capable to localize the beacon positions. While this method yields reliable results (and high spatial resolution), drawbacks are its relative complexity, inflexibility and difficulty to repair in case of failure. Also, it is difficult to include "passive" elements of complex setups which are not connected to power/acquisition systems.

This paper presents pilot test results of an alternative approach for acoustic localization of telescope elements. The main difference consists in the absence of any active beacons on the telescope. Localization of telescope elements is carried out "silently" as a result of 3D acoustic imaging and is based on the broad band acoustic echo signals from the telescope elements.

Acoustic techniques are widely used underwater for inspections, imaging, obstacle avoidance, etc. Highly intelligent sonars, including 3D sonars, are available and applied for various practical purposes. Main practical applications are imaging of sea bottom and/or single underwater objects. For non-standard tasks, however, like 3D imaging of a large number of discrete objects closely placed in the water volume, problems like appearance of "phantom" images will lead to reduced imaging capability.

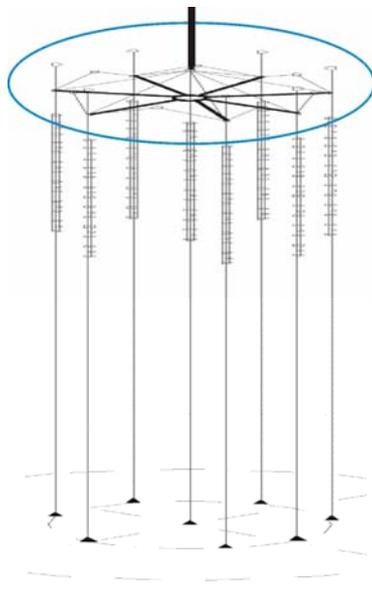

Fig. 1. Sketch of the Baikal Neutrino Telescope NT200:

In this paper we present first results of a 3D acoustic imaging system designed to cope with this specific tasks. Its key feature consists in localization of numerous discrete objects, placed closely to each other in the volume, and in suppression of the large number of "phantoms", appearing due to multipath propagation of probe and reflected signals, as well as due to spatial ambiguities of the processed images.

## 2. The experimental setup

The 3D acoustic imaging test has been carried out at the site of the Lake Baikal neutrino experiment [1,2] in early March, 2008 by EvoLogics GmbH, Berlin, jointly with the Baikal collaboration. The test was done before the yearly maintenance period of the telescope (during which the telescope NT200 is hauled up on the single carrying rope (see Fig.1.), serviced, and then re-deployed in early April. Due to the rotational degree of freedom during re-deployment, only the central string position is fixed; peripheral strings are situated on different positions for every year).

The main objective of this pilot 3D acoustic imaging test was to evaluate the possibilities for
(1) localization of reference elements (e.g. buoys) of the telescope by a temporary acoustic setup from the ice-surface, while the telescope stays in its standard working position (1100-1200 m),
(2) a future setup for a stationary acoustic imaging,
(3) future improvements of signal parameters.

Note, that (1) allows for an independent verification of the NT200 beacon acoustic system [2]; and is essential in case of its complete failure.

Fig.1 gives a sketch of the NT200 telescope - the central part of the NT200+ neutrino detector [1,2]. Seven peripheral strings and one central string are mounted on an umbrella-like frame with 21.5 m radius. The oval line in Fig. 1 indicates the telescope segment, which the main effort of this 3D acoustic imaging was concentrated on: the seven outer strings are suspended under cylindrical end buoys (~2.0 m height and ~1.0 m diameter), made of small aluminum spheres. Because of their large size, these string-buoys are good reflectors - their localization would determine the string positions. We note that two of the seven buoys have larger size (by ~ 50%); we also expect slight inclinations of some buoys – thus yielding different reflection strengths.

The acoustic antenna used in this test is made up of four acoustic transducers, spaced 50 m from each other and placed under the ice at 9.5m depth. A horizontal projection of the setup is shown in Fig.2. The antenna center is 91 m shifted horizontally from the center of the telescope, which is at ~1100m depth.

The sonification of the telescope was carried out by means of multiple-modulated wide-band signals, i.e. linear sweep-spread signals. The signals are similar to those used in underwater data transmission with hydro-acoustic modems of S2C technology (S2C - "sweep-spread carrier" communication technology) [3]. The sweep-spread signals had a linear frequency change from 10 to 22 kHz, with a length of 51.2 s. The signal level was 195 dB re 1 μPa, and the beam-width of the transmitted signal was 60 degrees.

The processing of the reflection signals from the telescope elements was similar to the procedure used by multi-static sonars [4].

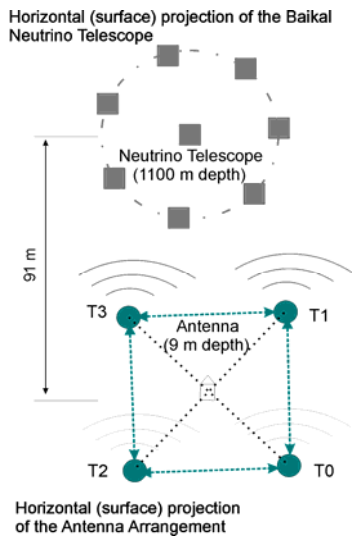

Fig. 2. Horizontal projection of the test setup: NT200 (1.1km depth) and acoustic antenna (9m below ice).

## 3. Experimental results

In Fig. 3 the result of the 3D acoustic imaging is visualized as a projection of the 3D picture onto the horizontal plane (see also fig.2). As can be seen, three objects were clearly found. They are visualized as evident (contrast) spots and labelled as "TC" (telescope center), "Obj3" and "Obj8" (buoys 3 and 8, respectively).

In general, the signals reflected from the telescope elements showed very low signal-to-noise ratios (SNR). After optimal filtration the SNR varied between 1 and 2 dB. We identified only 3 out of the 8 similar sized objects (buoys). The most reasonable explanation for this "loss" is the extremely low SNR, so that only the three objects with largest reflection strength (due to size and orientation) were detectable; the other buoys "disappeared" below noise level.

To verify the result (the positions of the resolved objects), the 3D acoustic image was overlaid with an independent measurement of the end buoys positions (by the regular acoustic localization system of NT200 [2]), as shown in Fig. 3 as white circles. We find good agreement. The accuracy of localization of the 3 objects is estimated as the width of the response main lobe at its half-height (standard method) - we find ~0.2 m along the beam and ~1.0 m transverse to the beam.

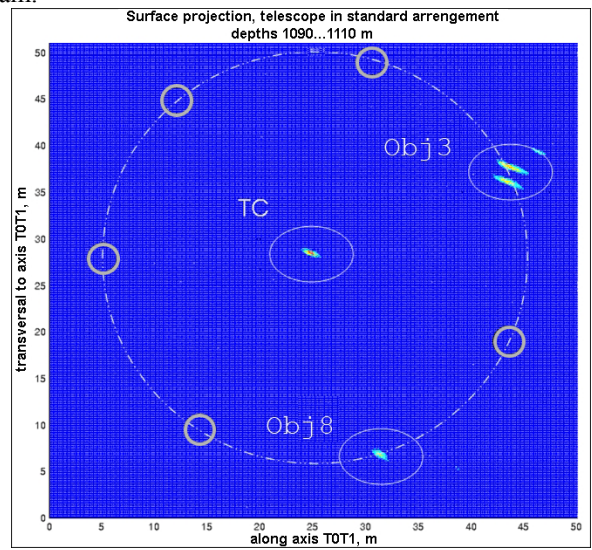

Fig. 3. Result of the 3D acoustic imaging test.

We mention, that the 3D imaging was obtained despite the complicating reverberating environment (strongly reflecting ice layer), as well as ambiguities due to the large number of objects with low-level reflection signals. Finally, it is worth noting that for the rotationally symmetric NT200 telescope the achieved determination of a few string positions is sufficient to calculate those of the remaining ones.

We estimate, that with increased energy content of the acoustic signals (longer duration), a gain of 3-6dB in SNR and thus localization of the remaining buoys should be feasible.

## 4. Perspective: BAN Antenna

Our pilot test has shown, that acoustic tomography is technically feasible; as demonstrated for the specific, ice-layer driven application for the Baikal telescope. For a stationary application, an alternative and more general approach can be considered - which is of interest also for other underwater telescopes (without a seasonal ice-cover): installation of acoustic antennas on the lake's bottom. This is sketched in figure 4. Three (or more) Bottom Acoustic Nodes (BANs) build an "antenna network". They enclose the underwater telescope (or parts of it, in case of multi-km scale arrays).

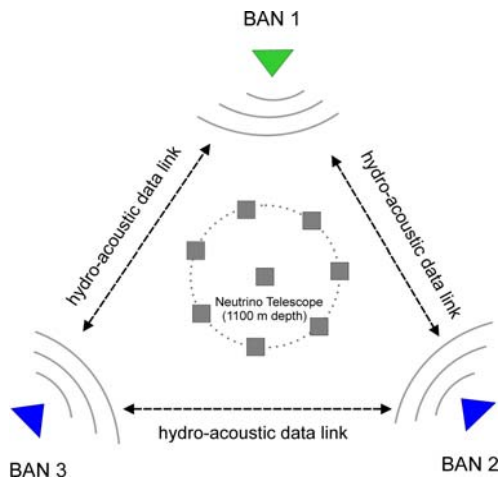

Fig.4: A bottom-based km-scale tomography system: 3 bottom acoustic nodes (BANs) enclose the underwater telescope.

Every BAN is made of an acoustic transducer (for sonification of the telescope), and a hydro-acoustic data modem of S2C technology [3] (for data exchange between the BANs). Using the inherent feature of S2C-modems to accurately evaluate their mutual distances, the positions of BANs and thus the antenna aperture can be precisely determined. The determination of absolute geographical coordinates for every BAN can be done via a geo-referenced surface modem (using another inherent S2C modem feature - the ultra-short base line ("USBL") ability).

This calibration is only carried out once from a ship.

Positioning results would be logged in one of the BANs and transmitted periodically (or on request) by the S2C-modem to the neutrino telescope acoustic modem (or to any other modem-equipped permanent or temporary unit; alternatively, also surface buoys equipped with radio/satellite capability can be used). BANs could work for a long time in autonomous mode (>1 yr), need only occasionally be recovered for battery replacement or re-charging.

Obviously, not only localization/positioning data can be transmitted by such an underwater acoustic network to the telescope (or to surface) – the acoustic modems can easily interface to a variety of devices collecting local underwater environmental data, seismic data, etc. - thus opening the road to a flexible, underwater array for "related science" [1].

## 5. Conclusions

This pilot study shows that accurate acoustic 3D imaging of medium-sized elements of underwater neutrino telescopes (e.g. buoys) is possible up to distances of km-scale, without using any active elements on the telescope.

We performed 3-dimensional imaging of key structural elements of the Baikal neutrino telescope main structure. The imaging system, installed directly below the ice-surface, located 3 major buoys at 1.1 km depth (longitudinal and transverse precision of 0.2 m and 1 m), under the presence of many close-by objects and with large reflections from the ice-surface. The method uses the linear sweep-spread signal based S2C-modem technology. Improvements of the currently quite low SNR are feasible.

For stationary application, autonomous antennas on the lake's bottom are suggested - using acoustic modems to connect with each other and a central data acquisition unit.

Acoustic imaging may be of interest for underwater neutrino telescopes of km3 scale and beyond, since it does not imply any active elements - thus being complementary to beacon-based systems. In particular, it can be applied to passive objects spread over large volumes, and/or in emergency situations.

**Acknowledgements**

The authors thank the Baikal collaboration for the opportunity to test the 3D hydro-acoustic imaging under realistic conditions.

**References**

[1] V. Aynutdinov et al.., The Baikal neutrino experiment – physics results and perspectives, these proceedings.

[2] I .Belolaptikov et al ., The Baikal underwater neutrino telescope: Design, perfromance and first results, Astropart. Phys. 7 (1997) 263.

[3] Kebkal K.G. and Bannasch R., ,Sweep-spread carrier for underwater communication over acoustic channels with strong multipath propagation, J. Acoust. Soc. Am., Vol.112, p. 2043.

[4] Coraluppi S. Multistatic Sonar Localization. *IEEE Journal of Oceanic Engineering*, vol. 31, no. 4, pp. 964-974.